\documentclass[twocolumn,showpacs,preprintnumbers,amsmath,amssymb]{revtex4}
\usepackage{graphicx}% Include figure files

\usepackage{dcolumn}% Align table columns on decimal point
\usepackage{bm}% bold math
\usepackage[latin1]{inputenc}

\begin{document}

%\preprint{APS/123-QED}

\title{Noise at a Fermi-edge singularity}% Force line breaks with \\

\author{N.~Maire$^{1}$}
\email{maire@nano.uni-hannover.de}
%\homepage{http://www.nano.uni-hannover.de}
\author{F.~Hohls$^{1}$}
\author{T.~L\"{u}dtke$^{1}$}
\author{K.~Pierz$^{2}$}
\author{R.~J.~Haug$^{1}$}
\affiliation{ $^{1}$Institut für Festkörperphysik, Universität
Hannover, Appelstr. 2, D-30167 Hannover, Germany\\
%$^{2}$Cavendish Laboratory, University of Cambridge Madingley
%Road, Cambridge CB3 0HE, UK\\
$^{2}$ Physikalisch-Technische Bundesanstalt, Bundesallee 100,
D-38116 Braunschweig, Germany }

\date{\today}% It is always \today, today,

\begin{abstract}
We present noise measurements of self-assembled InAs quantum dots
at high magnetic fields. In comparison to I-V characteristics at
zero magnetic field we notice a strong current overshoot which is
due to a Fermi-edge singularity. We observe an enhanced
suppression in the shot noise power simultaneous to the current
overshoot which is attributed to the electron-electron interaction
in the Fermi-edge singularity.
\end{abstract}

%Use showkeys class option if keyword

%%%%%%%%%%%%%%%%%%%%%%%% Einleitung %%%%%%%%%%%%%%%%%%%%%%%%%%%%%%%

\maketitle The measurement of shot noise provides information that
cannot be extracted from conductance measurements
alone~\cite{Blanter+Buettiker}. It has its origin in time
dependent fluctuations of the electrical current due to the
discreteness of the charge. For an uncorrelated flow of electrons
the shot noise power $S$ induced by individual tunneling events is
proportional to the stationary current $I$ and the absolute charge
of the electrons, $S=2eI$~\cite{Schottky}. Interactions between
the electrons e.g.\ Coulomb interaction or Pauli exclusion
principle can reduce the shot noise power~\cite{Li,Liu}. For zero
dimensional states, so called quantum dots, it has been shown both
theoretically and experimentally that the shot noise power $S$ is
suppressed down to half its normal value,
~\cite{Birk,Chen,Davies,Nauen'02}, $eI\leq S\leq 2eI$. Recently
deviations in the shot noise power have been reported due to
certain electron-electron interaction effects such as Kondo
effect~\cite{Meir,Lopez,Sela} or cotunneling~\cite{Thielmann}.

%%%%%%%%%%%%%%%%%%%%%%%% Zusammenfassung %%%%%%%%%%%%%%%%%%%%%%%%%%%%%

Motivated by these results we present temperature dependent noise
measurements of self-assembled InAs quantum dots under the
influence of a high magnetic field leading to another
electron-electron interaction effect, a so-called Fermi-edge
singularity effect. Its dominant feature is a strong overshoot in
the current at certain values of the bias voltage consistent with
other tunneling experiments at a localized
impurity~\cite{Geim_FES} or at InAs quantum
dots~\cite{Benedict_FES_PhysB'98,Isabella_FES_PRB'00}. We find
that this overshoot is accompanied by a surprising additional
suppression of the measured shot noise.

%%%%%%%%%%%%%%%%%%%%%%%% Probe %%%%%%%%%%%%%%%%%%%%%%%%%%%%%%%%%%%%%%%

The active part of the investigated sample consists of a
GaAs-AlAs-GaAs heterostructure. $N$-doped GaAs acts as 3d emitter
and collector. Situated inside the AlAs are 1.8 monolayers InAs.
Due to the Stranski-Krastanov growth mechanism InAs quantum dots
(QD) are formed. The lower and upper AlAs tunneling barriers are 4
and 6 nm thick, respectively. Since transmission electron
microscopy images show  that the QDs have a height of 2-3 nm the
effective thickness of the AlAs barrier on top of the QDs is
reduced to 3-4~nm.

%%%%%%%%%%%%%%%%%%%%%%%% Messaufbau %%%%%%%%%%%%%%%%%%%%%%%%%%%%%%%%%%%

The sample is inserted into a $^{3}$He system. This allows us to
reach temperatures down to $T = 300$ mK and magnetic fields up to
15 T. A DC bias is applied to the sample and the current is
amplified by a low noise current amplifier with a bandwidth of 20
kHz. The DC part is monitored by a voltmeter. A
fast-Fourier-transform analyzer measures the noise spectra.

\begin{figure}[t]
\includegraphics[scale=0.42]{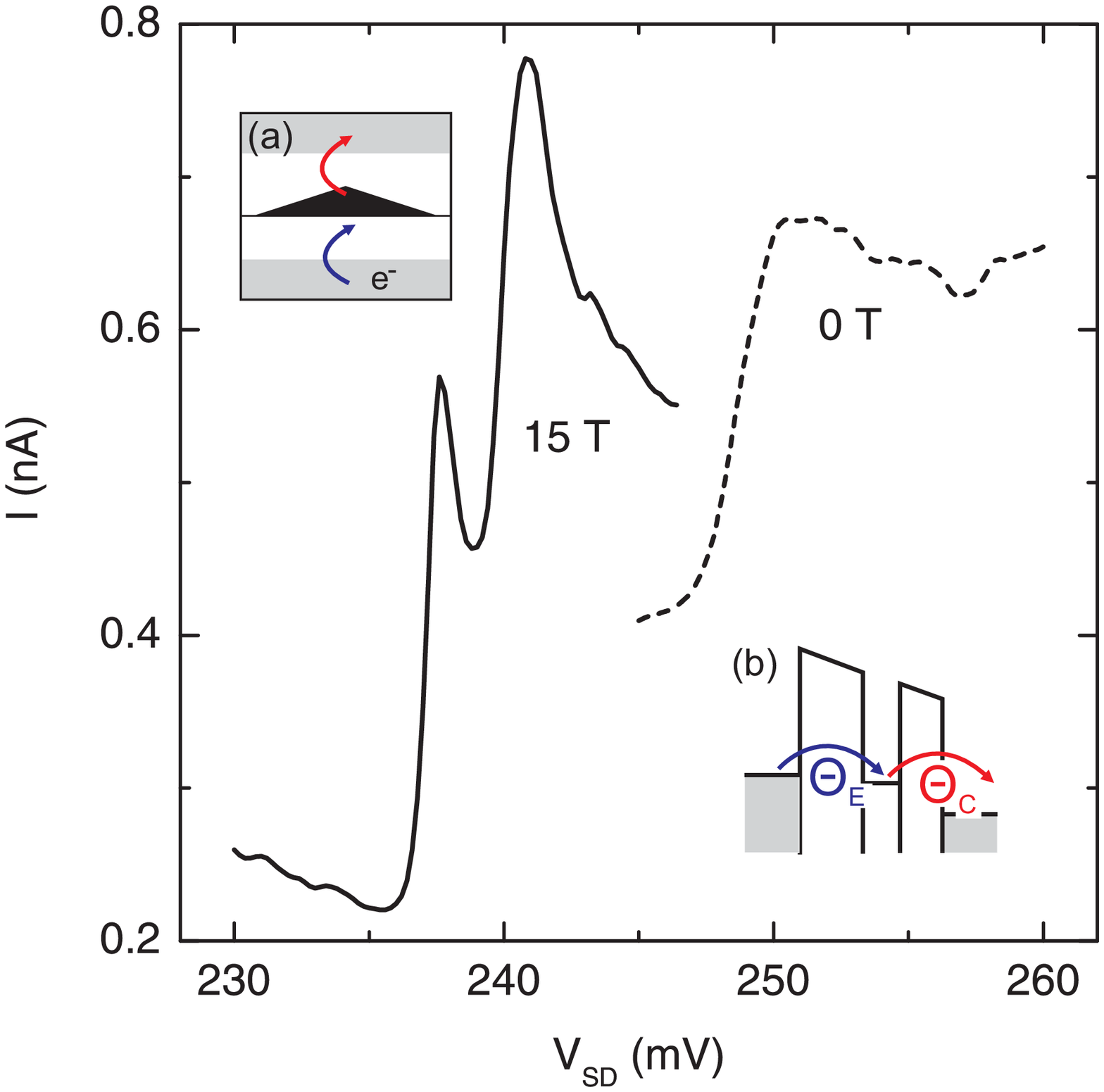}
\caption{\label{fig:eins} I-V characteristic of a current
step at $0$ T (dashed line) and at $15$~T (solid line).\\
Inset (a): Schematic of the tunneling direction of the electrons
through
the pyramidal shaped quantum dots.\\
Inset (b): Schematic of the corresponding band diagram when
resonant tunneling condition is achieved.}
\end{figure}

%%%%%%%%%%%%%%%%%%%%%%%% I/V %%%%%%%%%%%%%%%%%%%%%%%%%%%%%%%%%%%%%%%%%%%

The I-V characteristic of the sample shows distinct steps. These
steps correspond to resonant tunneling through individual quantum
dots~\cite{Isabella_APL'03}. It should be noted that at zero bias
the ground state energy of the quantum dots is well above the
Fermi energy in the leads~\cite{Nauen'02}. Resonant tunneling only
sets in when the applied bias voltage is sufficiently large to
bring the quantum dots' ground state energy on par with the
emitter Fermi edge. One of these steps due to resonant tunneling
is shown in Fig.~\ref{fig:eins} (dashed line). The applied
positive bias voltage corresponds to the tunneling direction
depicted in the inset (a). The electrons first tunnel through the
4 nm thick barrier into the dot and leave it through the thinner
3-4 nm barrier. So we expect the tunneling rate $\Theta_{C}$ out
of the dot into the collector to be higher than the emitter rate
$\Theta_{E}$ out of the emitter into the dot. We point out the
fact that transport measurements at a sample from the same wafer
structure also showed characteristics of a much higher collector
tunneling rate $\Theta_{C}$\cite{Nauen'04}. The fact that we
observe strong fluctuations in the I-V characteristic stemming
from fluctuations of the local density of states in the
emitter~\cite{Schmidt_LDOS} strengthens the expectation of
asymmetric tunneling rates with a much higher collector tunneling
rate $\Theta_{C}$. A schematic view of the corresponding band
structure is shown in inset (b).

%%%%%%%%%%%%%%%%%%%%%%%% Magnetfeld %%%%%%%%%%%%%%%%%%%%%%%%%%%%%%%%%%%

Also shown is the same current step at a high magnetic field of
15~T (solid line). The features discussed in the following were
also seen down to magnetic fields of $\approx 12$~T, 15~T was
chosen for the most distinct characteristics. \par At this large
magnetic field the resonance is shifted slightly to lower bias
voltages due to the fact that the emitter electrons have been
redistributed into the lowest Landau level leading to a lower
emitter Fermi energy $E_{F}$. It has split into two peaks due to
Zeeman splitting of the ground state of the quantum dot.

%%%%%%%%%%%%%%%%%%%%%%%% Fermi-edge singularity %%%%%%%%%%%%%%%%%%%%%%%%%

\begin{figure}[t]
\includegraphics[scale=0.42]{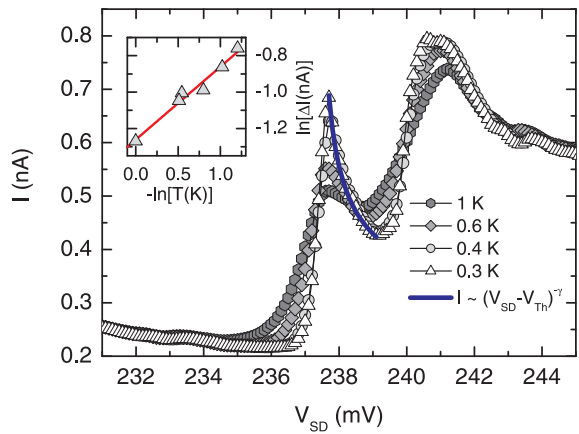}
\caption{\label{fig:zwei} I-V characteristic of the current step
at $B=15$ T for five different temperatures. Thick line: Fit using
Eq.~(\ref{eqn:zwei}).\\
Inset: Current step heights $\Delta I$ for different temperatures
(symbols) and corresponding fit using Eq.~(\ref{eqn:eins})
(line).}
\end{figure}

The more interesting feature is the strong peak like current
overshoot at the steps. The absolute current at $15$~T doubles
compared to $0$~T. The likely origin of this overshoot is the
Fermi-edge singularity effect which is an interaction effect
between the localized electron on the dot and the electrons near
the Fermi edge of the emitter. One signature of the Fermi-edge
singularity is a temperature dependence of the current step height
$\Delta I$ following the equation~\cite{Isabella_FES_PRB'00}
\begin{equation}
\label{eqn:eins} \ln(\Delta I)\sim -\gamma \cdot ln(T).
\end{equation}
In Fig.~\ref{fig:zwei} the current step at $15$~T is shown for
five different temperatures ranging from $T = 0.3$~K to $T = 1$~K.
The temperature dependent height of the peaks is clearly visible.
In the inset of Fig.~\ref{fig:zwei} a plot of $\ln(\Delta I)$ of
the first peak at $\approx 238$~mV vs.\ $-\ln(T)$ is shown. A fit
using Eq.~(\ref{eqn:eins}) yields $\gamma = 0.40 \pm 0.03$. For
similar measurements of a Fermi-edge singularity at InAs quantum
dots $\gamma = 0.43$ was obtained~\cite{Isabella_FES_PRB'00}.

%%%%%%%%%%%%%%%%%%%%%%%% FES I/V %%%%%%%%%%%%%%%%%%%%%%%%%%%%%%%%%%%

Another characteristic attribute of a Fermi-edge singularity
effect is a voltage dependence of the current given
by~\cite{Geim_FES}
\begin{equation}
\label{eqn:zwei} I \sim (V_{SD}-V_{Th})^{-\gamma}
\end{equation}
$V_{Th}$ corresponds to the bias voltage at which the Fermi energy
of the emitter is in resonance with the ground state of the dot.
From the I-V characteristics for different temperatures we
determine $V_{Th}=237.4$~mV. A fit of the 0.3~K data using
Eq.~(\ref{eqn:zwei}) with fixed $V_{Th}$ yields $\gamma = 0.47 \pm
0.01$ in fair agreement with the exponent determined above. This
fit is shown by the thick line in Fig.~\ref{fig:zwei} and matches
the measurement well. We conclude that this current overshoot is
indeed caused by a Fermi-edge singularity effect. A more
sophisticated analysis of the Fermi-edge singularity effect at a
current step can be found in Ref.~\cite{Frahm_FES_'06}
\begin{figure}[t]
\includegraphics[scale=0.42]{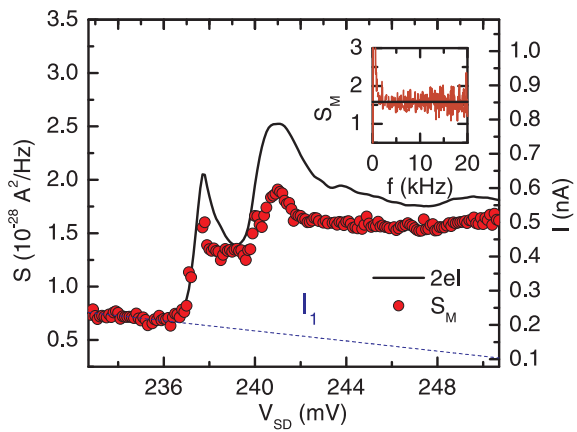}
\caption{\label{fig:drei} Measured shot noise power $S_{M}$
(filled dots, left axis) and current $I$ (solid line, right axis).
The scale on the right axis was chosen in such a way that the
solid line corresponds on the left axis to the shot noise power of
a single tunneling barrier, $S=2eI$. Also shown is the linear
fitted background current $I_{1}$ (dashed line, right axis).
\\
Inset: shot noise spectrum for $V_{SD}=244$~mV.}
\end{figure}

%%%%%%%%%%%%%%%%%%%%%%%% Rauschen %%%%%%%%%%%%%%%%%%%%%%%%%%%%%%%%%%

We will now analyze the noise characteristic of this particular
electron-electron interaction effect: At high frequencies the
measured noise power density is frequency independent as expected
for shot noise, while at low frequencies additional $1/f$ noise
appears. To remove the $1/f$ part, a $A/f + S_{M}$ fit is carried
out, $S_{M}$ being the resulting average shot noise power. A
sample spectrum is shown in the right inset of
Fig.~\ref{fig:drei}. For high differential source conductance we
additionally have to account for the input voltage noise of the
amplifier which adds a term $B/f^{\beta}$ with $\beta \approx 2$.
We include this term into the fit for the steep risers at $V_{SD}
\approx 237$~mV and $V_{SD} \approx 240$~mV. $S_{M}$ is shown in
Fig.~\ref{fig:drei} by the filled dots. Comparing $S_{M}$ to the
full shot noise $S=2eI$ of single barrier tunneling we find the
expected suppression of shot noise on resonance.

%%%%%%%%%%%%%%%%%%%%%%%% Rauschen semiklassisch %%%%%%%%%%%%%%%%%%%%%%

In a semiclassical picture the suppression of shot noise has its
origin in electron interaction effects such as Coulomb blockade
and Pauli exclusion principle. An emitter electron cannot enter
the dot when it is already occupied by another electron. This
anticorrelation then leads to the afore mentioned suppressed shot
noise compared to a single tunneling barrier. To better
characterize the degree of shot noise suppression the Fano factor
$\alpha = S_{M}/2eI$ is introduced. For zero temperature the Fano
factor $\alpha$ for a single ground state can be described
by~\cite{Chen}
\begin{equation}
\label{eqn:drei} \alpha = 1 -
\frac{2\Theta_{E}\Theta_{C}}{(\Theta_{E}+\Theta_{C})^{2}}
\end{equation}
with the emitter-dot tunneling rate $\Theta_{E}$ and collector-dot
tunneling rate $\Theta_{C}$. For a quantum dot $\alpha$ is
expected to be in the range of 0.5 to 1, 0.5 for symmetrical
barriers ($\Theta_{E}= \Theta_{C}$) and close to 1 for very
asymmetric barriers.
\begin{figure}[t]
\includegraphics[scale=0.42]{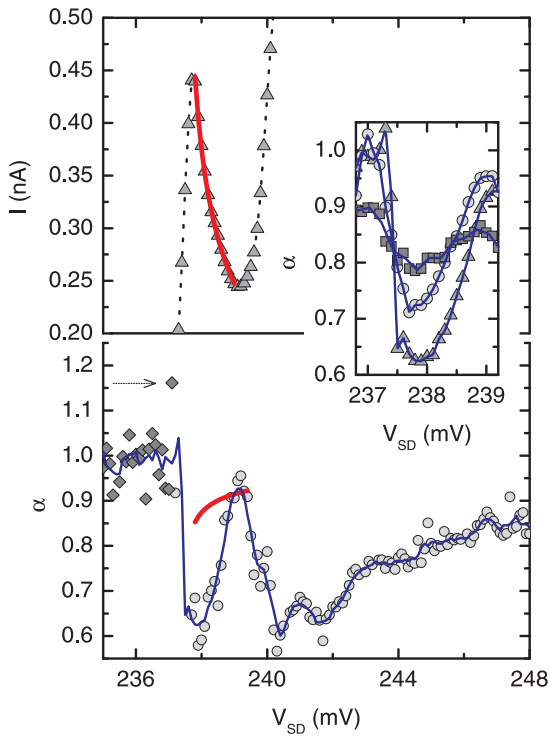}
\caption{\label{fig:vier}Fano factor $\alpha$ (left axis) at 0.4 K
using $\alpha=S_{M}/2eI$ (diamonds) and $\alpha_{1}$ (see text,
circles). The drawn through line is a boxcar average. The thick
solid line is a test of eq.~(\ref{eqn:drei}) (see text). Also
shown is the corresponding current $I$ (upper panel) and its fit
using eq.~(\ref{eqn:vier}) (see text, solid line).\\Inset: boxcar
averages of the Fano factor for 0.4 (triangles), 0.6 (circles) and
1 K (squares). The drawn through lines act as guides to the eye.}
\end{figure}

%%%%%%%%%%%%%%%%%%%%%%%% Fano Faktor %%%%%%%%%%%%%%%%%%%%%%%%%%%%%%%%

In Fig.~\ref{fig:vier} the measured Fano factor is shown. Below
237~mV $\alpha \approx 1$ is observed. This can be explained in
the following way: At $\approx 220$~mV resonant tunneling through
another quantum dot sets in. For resonant transport at voltages
sufficiently far away from the onset voltage a Fano factor of
$\alpha \approx 1$ is expected~\cite{Nauen'04}. We see an enhanced
shot noise ($\alpha = 1.16$, see arrow in Fig.~\ref{fig:vier})
just at the beginning of the current step at 237.1~mV. The origin
of this overshoot is unclear but we observe it consistently for
each measurement. With increasing temperatures the overshoot gets
less until it has completely vanished at 1~K. Super Poissonian
noise has also been recently observed at different quantum dot
systems~\cite{DD_Barthold'06,FCS_Gustavsson}.

%%%%%%%%%%%%%%%%%%%%%%%% Alpha ohne Ensemble %%%%%%%%%%%%%%%%%%%%%%%%%

To extract the Fano factor originating only from the quantum dot
participating in the resonant tunneling process with the
Fermi-edge singularity effect we have to subtract the influence of
the other afore mentioned dot being resonant at $\approx 220$~mV.
The contribution $\alpha_{1,2}$ of the two quantum dots to the
Fano factor is given by their fractions $I_{1,2}$ of the overall
current $I$~\cite{Nauen'02,Kiesslich_QDArray_PSS'03}: A linear fit
of the current $I_{1}$ of the dot with resonance at $220$~mV is
depicted by the dashed line in Fig.~\ref{fig:drei}. If we now
calculate $\alpha_{2}$ with an $\alpha_{1}=1$ and the linear
fitted current $I_{1}$ we get the Fano factor given by the circles
in Fig.~\ref{fig:vier}. The line is a 3 pt. boxcar average of
these data.

%%%%%%%%%%%%%%%%%%%%%%%% Alpha Diskussion %%%%%%%%%%%%%%%%%%%%%%%%%%%%%%%

As can be seen the Fano factor $\alpha$ exhibits a very sharp dip
at the onset of the current step ($V_{SD}=237$~mV) with a maximum
suppression at the voltage position of the current peak. For
further increased voltage we observe a strong rise of $\alpha$ in
parallel to the large decrease in current. The pattern is repeated
when the upper Zeeman level of the quantum dot comes into
resonance at $V_{SD}=240$~mV. We will restrict the following
discussion of the Fano factor to the range $V_{SD} \leq 239$~mV
where the second Zeeman level can be safely ignored as only a very
small fraction of the current is carried by this second level.

The initial drop of $\alpha$ can be qualitatively explained
following Ref.~\cite{Nauen'04}: At the onset of resonant tunneling
only the highest energetic emitter electrons in the tail of the
emitter Fermi distribution function $f_{E}$ can participate in a
resonant tunneling process and the effective tunneling rate can be
written as $\Theta_{E} = \Theta^{0}_{E} f_{E}$ with $f_{E} \ll 1$
and thus $\Theta_{E} \ll \Theta_{C}$. Therefore we start with
$\alpha \approx 1$. With increasing bias voltage the energy level
of the dot is shifted downwards with respect to the Fermi level of
the emitter and $f_E = 0 \rightarrow 1$. $\Theta_{E}$ increases
accordingly, leading to a rise in the current and with
Eq.~(\ref{eqn:drei}) to a decrease of the Fano factor.

%%%%%%%%%%%%%%%%%%%%%%%% starker Anstieg von Alpha %%%%%%%%%%%%%%%%%%%%%%

In our previous measurements of similar devices at $B=0$ we have
shown that the initial drop of the Fano factor is followed by a
gentle rise over several ten mV which mirrors the density of
states in the emitter~\cite{Nauen'04}. In our measurement here for
15~T we observe a change of $\alpha$ on a totally different
voltage scale: The Fano factor rises rapidly from $\alpha = 0.58$
at 237.9~mV to $\alpha = 0.91$ at 238.9~mV. Both Fano factor and
current change dramatically within a voltage range of only 1~mV!

We indeed anticipate a rapid change near the Fermi energy in the
presence of a Fermi-edge singularity. We will now try whether the
measured current $I$ and Fano factor $\alpha$ can be modeled by
the introduction of an interaction enhanced emitter tunneling rate
$\Theta_{E}$ into the semiclassical relations for $I$ and
$\alpha$. In the zero temperature limit $\Theta_{E}$ is predicted
to follow~\cite{Matveev}:
\begin{equation}
\label{eqn:vier} \Theta_{E}(V_{SD}) =
\Theta^{0}_{E}\left(\frac{D}{e(V_{SD}-V_{Th})}\right)^{\gamma}
\end{equation}
for a voltage near but not too close to the threshold voltage
$V_{Th}$. Following Ref.~\cite{Matveev} the increase of the
current peak can then be described by
\begin{equation}
\label{eqn:fuenf}
I=e\frac{\Theta_{C}\Theta_{E}(V_{SD})}{\Theta_{C}+\Theta_{E}(V_{SD})}
\end{equation}
Due to the high bias the change of the collector tunneling rate
$\Theta_{C}$ is negligible in the range of
interest~\cite{Nauen'04}, and we determine it far away from the
resonances where the effect of the Fermi-edge singularity is
negligible [$\Theta_{C}= 3.4\cdot 10^{10}$ s$^{-1}$ at
$V_{SD}=248$~mV]. Inserting $\Theta_{C}$ in Eq.~(\ref{eqn:fuenf})
and fitting the first current peak [$V_{Th}=237.4$~mV fixed;
$\gamma = 0.45$ and $\Theta_{E}^{0}(D/e)^{\gamma} = 8.4\cdot
10^{7}$~V$^{\gamma}$/s fitted] gives sound agreement with the
measurement as depicted in Fig.~\ref{fig:vier} by the thick solid
line in the upper panel.

%%%%%%%%%%%%%%%%%%%%%%%% Test der Theorie %%%%%%%%%%%%%%%%%%%%%%%%%%%%%%%
We can now insert the above determined interaction enhanced and
voltage dependent tunneling rate $\Theta_{E}(V_{SD})$ into the
semiclassical equation Eq.~(\ref{fig:drei}). The resulting
prediction for $\alpha$ is depicted by the thick solid line in the
lower panel of Fig.~\ref{fig:vier}. The values calculated at some
distance to the Fermi-edge singularity where the interaction
effects are weak are in reasonable agreement with our measurement
(e.g.~$\alpha = 0.92$ at $V_{SD}=239.4$~mV). However, near to the
Fermi-edge singularity we observe a significant discrepancy: The
calculated shot noise suppression of $\alpha = 0.86$ at
$V_{SD}=237.9$~mV falls way short of the measured strong
suppression $\alpha=0.58$; the measured change of $\alpha$ is much
more rapid than the calculated one.

Thus we find that Eq.~(\ref{fig:drei}) is not applicable near the
Fermi edge singularity. The Fano factor cannot be accounted for by
just inserting an interaction enhanced tunneling rate $\Theta_E$
into a relation that was deduced in a sequential tunneling
picture. Instead we observe a strongly reduced Fano factor hinting
onto additional anticorrelations of the tunneling events due to
the interaction between lead and dot at the Fermi edge
singularity.

%%%%%%%%%%%%%%%%%%%%%%%% Temperatur %%%%%%%%%%%%%%%%%%%%%%%%%%%%

The strong impact of the Fermi-edge singularity on the shot noise
is further confirmed by the influence of temperature; the inset of
Fig.~\ref{fig:vier} shows the temperature dependence of the Fano
factor near the resonance. We observe a much stronger temperature
dependence than expected when just using the changing Fermi
function $f_{E}$ in the effective tunneling rate $\Theta_{E} =
\Theta^{0}_{E} f_{E}$ at the step edge~\cite{Nauen'04}.

%%%%%%%%%%%%%%%%%%%%%%%% Bunching, Kondo %%%%%%%%%%%%%%%%%%%%%%%%%%%%

We conclude that the current noise in the regime of the Fermi-edge
singularity reveals the coherent nature of the interaction between
the emitter and the dot. Only a theory that accounts for this
interaction will be able to describe the shot noise near the
singularity. The relevance of interactions between the lead and
the dot were pointed out for quantum dots in the regime of large
tunnel coupling where the Kondo effect is observed. A number of
{\sl theoretical} papers~\cite{Meir,Lopez,Sela} emphasized the
importance of noise measurements to probe the Kondo regime, but
due to the difficulty of the measurement experimental data in this
regime are still missing. The need to go beyond sequential
tunneling in a regime of lower coupling was demonstrated in {\sl
calculations} for an Anderson-impurity model with finite spin
splitting~\cite{Thielmann} and now awaits experimental
verification. For both the above mentioned Kondo and cotunneling
regime the spin degree of freedom on the dot was essential. In
contrast our experiment demonstrates the relevance of interaction
for a single level system with only one spin species involved.
Thus beside {\sl experimentally} demonstrating the relevance of
shot noise to examine the dot-lead interaction we also reveal the
importance of current correlations in a new regime.

%%%%%%%%%%%%%%%%%%%%%%%% Zusammenfassung %%%%%%%%%%%%%%%%%%%%%%%%%%%%

In summary we have measured the shot noise at a Fermi-edge
singularity. We have observed strong shot noise suppression which
we attribute to strong interaction between lead and dot at the
Fermi-edge singularity.

%%%%%%%%%%%%%%%%%%%%%%%% BMBF %%%%%%%%%%%%%%%%%%%%%%%%%%%%%%%%%%%%%%%

We acknowledge financial support from BMBF via the program
"NanoQUIT".


\begin{thebibliography}{99}
\bibitem{Blanter+Buettiker}
    \bibinfo{author}{Y. M. Blanter} and
    \bibinfo{author}{M. Büttiker},
    \bibinfo{journal}{Phys. Rep.}
    \textbf{\bibinfo{volume}{336}},
    \bibinfo{pages}{1}
    (\bibinfo{year}{2000}).

\bibitem{Schottky}
    \bibinfo{author}{W. Schottky},
    \bibinfo{journal}{Ann. Phys. (Leipzig)}
    \textbf{\bibinfo{volume}{57}},
    \bibinfo{pages}{541}
    (\bibinfo{year}{1918}).

\bibitem{Li}
    \bibinfo{author}{Y. P. Li},
    \bibinfo{author}{A. Zaslavsky},
    \bibinfo{author}{D. C. Tsui},
    \bibinfo{author}{M. Santos}, and
    \bibinfo{author}{M. Shayegan},
    \bibinfo{journal}{Phys. Rev. B}
    \textbf{\bibinfo{volume}{41}},
    \bibinfo{pages}{8388}
    (\bibinfo{year}{1990}).

\bibitem{Liu}
    \bibinfo{author}{H. C. Liu},
    \bibinfo{author}{J. Li},
    \bibinfo{author}{G. C. Aers},
    \bibinfo{author}{C. R. Leavens},
    \bibinfo{author}{M. Buchanan}, and
    \bibinfo{author}{Z. R. Wasilewski},
    \bibinfo{journal}{Phys. Rev. B}
    \textbf{\bibinfo{volume}{51}},
    \bibinfo{pages}{5116}
    (\bibinfo{year}{1995}).

\bibitem{Birk}
    \bibinfo{author}{H. Birk},
    \bibinfo{author}{M. J. M. de Jong}, and
    \bibinfo{author}{C. Sch\"{o}nenberger},
    \bibinfo{journal}{Phys. Rev. Lett.}
    \textbf{\bibinfo{volume}{75}},
    \bibinfo{pages}{1610}
    (\bibinfo{year}{1995}).

\bibitem{Chen}
    \bibinfo{author}{L. Y. Chen} and
    \bibinfo{author}{C. S. Ting},
    \bibinfo{journal}{Phys. Rev. B}
    \textbf{\bibinfo{volume}{43}},
    \bibinfo{pages}{4534}
    (\bibinfo{year}{1991}).

\bibitem{Davies}
    \bibinfo{author}{J. H. Davies},
    \bibinfo{author}{P. Hyldgaard},
    \bibinfo{author}{S. Hershfield}, and
    \bibinfo{author}{J. W. Wilkins},
    \bibinfo{journal}{Phys. Rev. B}
    \textbf{\bibinfo{volume}{46}},
    \bibinfo{pages}{9620}
    (\bibinfo{year}{1992}).

\bibitem{Nauen'02}
    \bibinfo{author}{A. Nauen},
    \bibinfo{author}{I. Hapke-Wurst},
    \bibinfo{author}{F. Hohls},
    \bibinfo{author}{U. Zeitler},
    \bibinfo{author}{R. J. Haug}, and
    \bibinfo{author}{K. Pierz},
    \bibinfo{journal}{Phys. Rev. B}
    \textbf{\bibinfo{volume}{66}},
    \bibinfo{pages}{161303(R)}
    (\bibinfo{year}{2002}).

\bibitem{Meir}
    \bibinfo{author}{Y. Meir} and
    \bibinfo{author}{A. Golub},
    \bibinfo{journal}{Phys. Rev. Lett.}
    \textbf{\bibinfo{volume}{88}},
    \bibinfo{pages}{116802}
    (\bibinfo{year}{2002}).

\bibitem{Lopez}
    \bibinfo{author}{R. Lopez} and
    \bibinfo{author}{D. Sanchez},
    \bibinfo{journal}{Phys. Rev. Lett.}
    \textbf{\bibinfo{volume}{90}},
    \bibinfo{pages}{116602}
    (\bibinfo{year}{2003}).

\bibitem{Sela}
    \bibinfo{author}{E. Sela},
    \bibinfo{author}{Y. Oreg},
    \bibinfo{author}{F. von Oppen}, and
    \bibinfo{author}{J. Koch},
    \bibinfo{journal}{Phys. Rev. Lett.}
    \textbf{\bibinfo{volume}{97}},
    \bibinfo{pages}{086601}
    (\bibinfo{year}{2006}).

\bibitem{Thielmann}
    \bibinfo{author}{A. Thielmann},
    \bibinfo{author}{M. H. Hettler},
    \bibinfo{author}{J. K\"onig}, and
    \bibinfo{author}{G. Sch\"on},
    \bibinfo{journal}{Phys. Rev. Lett.}
    \textbf{\bibinfo{volume}{95}},
    \bibinfo{pages}{146806}
    (\bibinfo{year}{2005}).

\bibitem{Geim_FES}
    \bibinfo{author}{A. K. Geim},
    \bibinfo{author}{P. C. Main},
    \bibinfo{author}{N. La Scala Jr},
    \bibinfo{author}{L. Eaves},
    \bibinfo{author}{T. J. Foster},
    \bibinfo{author}{P. H. Beton},
    \bibinfo{author}{J. W. Sakai},
    \bibinfo{author}{F. W. Sheard},
    \bibinfo{author}{M. Henini},
    \bibinfo{author}{G. Hill}, and
    \bibinfo{author}{M. A. Pate},
    \bibinfo{journal}{Phys. Rev. Lett.}
    \textbf{\bibinfo{volume}{72}},
    \bibinfo{pages}{2061}
    (\bibinfo{year}{1994}).

\bibitem{Benedict_FES_PhysB'98}
    \bibinfo{author}{K. A. Benedict},
    \bibinfo{author}{A. S. G. Thornton},
    \bibinfo{author}{T. Ihn},
    \bibinfo{author}{P. C. Main},
    \bibinfo{author}{L. Eaves}, and
    \bibinfo{author}{M. Henini},
    \bibinfo{journal}{Physica B}
    \textbf{\bibinfo{volume}{256-258}},
    \bibinfo{pages}{519}
    (\bibinfo{year}{1998}).

\bibitem{Isabella_FES_PRB'00}
    \bibinfo{author}{I. Hapke-Wurst},
    \bibinfo{author}{U. Zeitler},
    \bibinfo{author}{H. Frahm},
    \bibinfo{author}{A. G. M. Jansen},
    \bibinfo{author}{R. J. Haug}, and
    \bibinfo{author}{K. Pierz},
    \bibinfo{journal}{Phys. Rev. B}
    \textbf{\bibinfo{volume}{62}},
    \bibinfo{pages}{12621}
    (\bibinfo{year}{2000}).


\bibitem{Isabella_APL'03}
    \bibinfo{author}{I. Hapke-Wurst},
    \bibinfo{author}{U. Zeitler},
    \bibinfo{author}{U. F. Keyser},
    \bibinfo{author}{R. J. Haug},
    \bibinfo{author}{K. Pierz}, and
    \bibinfo{author}{Z. Ma},
    \bibinfo{journal}{Appl. Phys. Lett.}
    \textbf{\bibinfo{volume}{82}},
    \bibinfo{pages}{1209}
    (\bibinfo{year}{2003}).

\bibitem{Schmidt_LDOS}
    \bibinfo{author}{T. Schmidt},
    \bibinfo{author}{R. J. Haug},
    \bibinfo{author}{V. I. Fal'ko},
    \bibinfo{author}{K. v. Klitzing},
    \bibinfo{author}{A. Förster}, and
    \bibinfo{author}{H. Lüth},
    \bibinfo{journal}{Europhys. Lett.}
    \textbf{\bibinfo{volume}{36}},
    \bibinfo{pages}{61}
    (\bibinfo{year}{1996}).

\bibitem{Frahm_FES_'06}
    \bibinfo{author}{H. Frahm},
    \bibinfo{author}{C. von Zobeltitz},
    \bibinfo{author}{N. Maire}, and
    \bibinfo{author}{R. J. Haug},
    \bibinfo{journal}{Phys. Rev. B}.
    \textbf{\bibinfo{volume}{74}}
    \bibinfo{pages}{035329}
    \bibinfo{year}{2006}.

\bibitem{Nauen'04}
    \bibinfo{author}{A. Nauen},
    \bibinfo{author}{F. Hohls},
    \bibinfo{author}{N. Maire},
    \bibinfo{author}{K. Pierz}, and
    \bibinfo{author}{R. J. Haug},
    \bibinfo{journal}{Phys. Rev. B}
    \textbf{\bibinfo{volume}{70}},
    \bibinfo{pages}{033305}
    (\bibinfo{year}{2004}).

\bibitem{DD_Barthold'06}
    \bibinfo{author}{P. Barthold},
    \bibinfo{author}{F. Hohls},
    \bibinfo{author}{N. Maire},
    \bibinfo{author}{K. Pierz}, and
    \bibinfo{author}{R. J. Haug},
    \bibinfo{journal}{Phys. Rev. Lett.}
    \textbf{\bibinfo{volume}{96}},
    \bibinfo{pages}{246804}
    (\bibinfo{year}{2006}).

\bibitem{FCS_Gustavsson}
    \bibinfo{author}{S. Gustavsson},
    \bibinfo{author}{R. Leturcq},
    \bibinfo{author}{B. Simovi\u{c}},
    \bibinfo{author}{R. Schleser},
    \bibinfo{author}{P. Studerus},
    \bibinfo{author}{T. Ihn},
    \bibinfo{author}{K. Ensslin},
    \bibinfo{author}{D. C. Driscoll}, and
    \bibinfo{author}{A. C. Gossard},
    \bibinfo{journal}{cond-mat/0605365}.
    %\textbf{\bibinfo{volume}{96}},
    %\bibinfo{pages}{076605}
    %(\bibinfo{year}{2006}).

\bibitem{Kiesslich_QDArray_PSS'03}
    \bibinfo{author}{G. Kiesslich},
    \bibinfo{author}{A. Wacker},
    \bibinfo{author}{E. Schöll},
    \bibinfo{author}{A. Nauen},
    \bibinfo{author}{F. Hohls}, and
    \bibinfo{author}{R. J. Haug},
    \bibinfo{journal}{phys. stat. sol. (c)}
    \textbf{\bibinfo{volume}{0}},
    \bibinfo{pages}{1293}
    (\bibinfo{year}{2003}).

\bibitem{Matveev}
    \bibinfo{author}{K. A. Matveev} and
    \bibinfo{author}{A. I. Larkin},
    \bibinfo{journal}{Phys. Rev. B}
    \textbf{\bibinfo{volume}{46}},
    \bibinfo{pages}{15337}
    (\bibinfo{year}{1992}).

%\bibitem{Belzig}
%    \bibinfo{author}{W. Belzig},
%    \bibinfo{journal}{Phys. Rev. B}
%    \textbf{\bibinfo{volume}{71}},
%    \bibinfo{pages}{161301(R)}
%    (\bibinfo{year}{2005}).

\end{thebibliography}
\end{document}